\documentclass[sn-mathphys,Numbered]{sn-jnl}

\usepackage{graphicx}%
\usepackage{multirow}%
\usepackage{amsmath,amssymb,amsfonts}%
\usepackage{amsthm}%
\usepackage{mathrsfs}%
\usepackage[title]{appendix}%
\usepackage{xcolor}%
\usepackage{textcomp}%
\usepackage{manyfoot}%
\usepackage{booktabs}%

\begin{document}

\title[Article Title]{Spectral study of X-ray sources in some galaxies recently observed by {\it Chandra}}


\author[1]{\fnm{Amom} \sur{Lanchenbi Chanu}}\email{lanchenbi.phd.phy@manipuruniv.ac.in}

\author*[1]{\fnm{Anoubam} \sur{Senorita Devi}}\email{senorita@manipuruniv.ac.in}

\affil[1]{\orgdiv{Department of Physics}, \orgname{Manipur University}, \orgaddress{\street{}, \city{Canchipur}, \postcode{795003}, \state{Manipur}, \country{India}}}

\abstract{With the aim to study the spectral properties of some X-ray sources from recently observed {\it Chandra} data, 9 galaxies which have been observed by {\it Chandra} ACIS-S during the year 2018 to 2022 have been considered for the present work. 27 sources with net source counts $ \ge$ 100 have been considered. The spectra of all the sources were fitted using two empirical models- an absorbed powerlaw and an absorbed disk blackbody. From their estimated bolometric luminosities, the 27 X-ray sources are categorized as 6 X-ray binaries (XRBs) and 21 Ultraluminous X-ray sources (ULXs). All the six XRBs are found to be in the spectrally hard state ($\Gamma \sim$ 1.52-2.29) which indeed may be due to thermal comptonization. Only one ULX - CXOUJ032251.2-370950 (X-5) was found to be spectrally soft while the remaining 20 ULXs were spectrally hard. The spectral parameters of X-5 with an inner disk temperature (kT$_{in}$) $\sim $ 0.5 keV and an estimated bolometric luminosity, L$_X \sim$ 3.26 $\times$ 10$^{39}$ erg s$^{-1} $ requires a black hole of mass,  M$_{BH} \sim$ 137.86$^{+66.62}_{-47.41}$ M$_\odot $ accreting at $  \sim$ 0.19  times its Eddington limit. 8 ULXs - X-4, X-8, X-9, X-10, X-11, X-12, X-20 and X-21 were found to be in the Extremely luminous X-ray sources (ELXs) regime with even their lower limit of luminosity $>$ 10$^{40}$ erg s$^{-1}$. Softening/Hardening of spectra with or without changes in the luminosity were also observed in some ULXs/ELXs. In the hard ELX, X-8, spectral softening with almost consistent luminosity was observed. While in the ULXs - X-20 and X-25 spectral softening with increasing luminosity was observed. However spectral hardening with increase in luminosity were observed in the ULXs - X-21 and X-26.}

\keywords{accretion, accretion disks; galaxies: general-X-rays:binaries}



\maketitle
\clearpage

\section{Introduction}\label{sec1}

Ultraluminous X-ray sources (ULXs) are defined as  point-like, non nuclear X-ray sources with their isotropic X-ray luminosity, $L_X \gtrsim 10^{39}$ erg s$^{-1}$ in the energy range 0.3-10.0 keV, which actually exceeds the isotropic Eddington luminosity for a standard black hole (BH) of 10 M$_\odot$ \cite {kaaret2017ultraluminous}. ULXs are rare and elusive objects that were primarily observed in external galaxies during their early exploration. Nevertheless, in recent times, there has been a  noteworthy shift as galactic pulsating ULXs (PULXs) are now being identified as well. Now a large number of ULXs, above $\approx 1840$ in count, has been identified, and their population has been comprehensively studied \cite {swartz2004ultraluminous} \cite {devi2007dependence} \cite {kaaret2017ultraluminous} \cite {earnshaw2019new} \cite {kovlakas2020census} \cite {walton2022multimission}. Since its discovery \cite {fabbiano1989x}, the understanding of the physical nature of ULXs, including the mass accretion rate and and the mass of the compact object harbored by ULXs have been the subject of controversy and ongoing debate.\\
Various models have been suggested to account for the high X-ray luminosities observed in ULXs, including (i) super-Eddington accretion on to stellar mass BHs with mass, M$_{BH} \approx $ 10M$_{\odot}$ \cite {mukherjee2015hard} \cite{middleton2015diagnosing} \cite{motch2014mass} \cite{liu2013puzzling} \cite{singha2019black}) or (ii) sub-Eddington accretion onto Intermediate mass black holes (IMBHs) with mass, M$_{BH}$ $\sim 10^2 - 10^5$ M$_{\odot}$ \cite{godet2012investigating} \cite{singha2019bimodal} or (iii) relativistic and geometric beaming from an anisotropic super-critical accretor  \cite {king2001ultraluminous} \cite{king2023ultraluminous}, where the accretor can either be a stellar mass black hole or a neutron star with mass $ \approx $ 1-2 M$_{\odot}$  \cite{sathyaprakash2019discovery}. 
Based on high quality XMM Newton ULX spectra, ULXs can be spectrally classified as - i) Supersoft, characterized by an inner disk temperature, kT$_{in}$ $\sim$ 0.1 keV as observed in the ULX, NGC 247 X-1 ii) Soft type, exhibiting an inner disk temperature, kT$_{in}$ $\sim$ 0.1-0.5 keV akin to the ULX NGC 5408 X-1 iii) Hard type, distinguished by an inner disk temperature kT$_{in}$ $\gtrsim$ 1.0 keV, exemplified by NGC 1313 X-1, and similar sources \cite {sutton2013bright},\cite{pinto2023ultra}. However, for most Chandra ULXs spectra, due to the relatively lower source count, the ULXs spectra are commonly classified into hard state (kT$_{in}$ $\gtrsim$ 1.0 keV) and soft states (kT$_{in}$ $\le $ 0.5 keV) only.\\
Previously, an extensive and detailed study of the X-ray spectra and variability features of certain ULXs has indicated their association with accreting compact objects in binary systems  \cite {urquhart2018multiband}. Moreover, ASCA X-ray spectral studies on numerous ULXs have provided evidence supporting the idea that ULXs exhibit characteristics consistent with accreting black holes. Subsequently, through the utilization of high-quality XMM-Newton data and data from advanced detectors like those on Chandra, a detailed investigation into the ULX spectra and variability of individual sources has been conducted. The results indicate that certain ULXs can exhibit different spectral states analogous to those observed in X-ray binaries \cite {kaaret2017ultraluminous}. Furthermore, spectral state transitions between two states have been reported in many ULXs, as exemplified by NGC 1313 X-1 \cite{feng2006spectral}. Indeed, all these findings point towards that some ULXs are quite closely related to black hole binaries. So, if ULX sources are indeed accreting compact objects, then their very high luminosity strongly suggests for the ULXs to harbor massive (stellar mass) black holes. Subsequent studies have also given the evidence for these ULXs to be powered by accretion on to stellar mass black holes such as that of ULX, X-6, in the nearby galaxy NGC 4258 M106 \cite {avdan2016x}, the ULXs in NGC 2276 which were also reported to be powered by stellar mass black holes \cite {singha2021spectral} etc.

Even though the overall population of ULXs are generally found to be dominated by super-Eddington accretors, there are still rare individual sources among the ULX population that are highly luminous ($L_X \ge 10^{40}$ erg s$^{-1}$) that fall within the Intermediate-Mass Black Hole (IMBH) regime. The importance of IMBH theory is that it provides the missing link between the stellar mass BHs and the super massive BHs \cite{ebisuzaki2001missing} \cite{singha2021spectral} . The formation of Intermediate-Mass Black Holes (IMBHs) and the mechanisms by which they maintain exceptionally high accretion rates remain inadequately understood. However, recent observations with LIGO has confirmed the detection of a binary black hole merger with total mass, $M_{BH} \sim 150 M_{\odot}$ \cite{abbott2020gw190521}. This discovery serves as evidence for the formation of Intermediate-Mass Black Holes (IMBHs). Two subgroups of ULXs - the Extremely luminous X-ray sources (ELXs) with $L_X \approx 10^{40}$ erg s$^{-1}$  and the Hyperluminous X-ray sources (HLXs) with even higher luminosities ($L_X $) exceeding  $ 10^{41} $erg s$^{-1}$ are considered prime candidates for harboring Intermediate-Mass Black Holes (IMBHs). This is because, in order to generate such high luminosities at a near-Eddington rate, BH masses around $10^2 - 10^5 M_\odot$ are required as central compact object \cite{sutton2012most}. The evidence for such BHs gains strong support from the discovery of the most luminous HLX (HLX-1) on the outskirts of the edge-on spiral galaxy ESO 243-49 with a peak luminosity of the order of $10^{42}$ erg s$^{-1}$ corresponding to a BH in the intermediate mass range, 3000 $M_\odot < M_{BH} < 3 \times 10^5 M_\odot$ \cite{godet2012investigating}. 
	
The latest breakthrough discovery for ULX model is the ULX pulsar. Pulsations were recently detected in many ULXs, including one galactic ULX  \cite {doroshenko2020first}which gives an evidence for some ULXs to harbor pulsars also.  In such cases, ULXs are  modelled as an accreting system in which the compact object is a neutron star with mass  $\sim$ 1 - 2 M$_{\odot}$ , accreting at extreme super-Eddington rates. Notably, NuSTAR observations of the starburst galaxy M82 marked the initial detection of ULX pulsations \cite{bachetti2014ultraluminous}, subsequently many other ULXs were confirmed to  present pulsations \cite{israel2017discovery} \cite{carpano2018discovery} \cite{sathyaprakash2019discovery} \cite{castillo2020discovery}.  The ULX - Swift J0243.6+6124 stands out as the first reported galactic pulsating ULX \cite {doroshenko2020first}. 	

There are also many other models proposing ULXs as supernova remnants \cite{mezcua2011compact}, accreting white dwarfs \cite{xu2012spitzer} and background AGNs \cite{dadina2013ultraluminous} . Consequently, ULXs appear to constitute an extremely non uniform group of objects, with some undergoing super-critical accretion and others accreting below the Eddington limit. Additionally, the accreting compact objects within ULXs seem to encompass neutron stars,  stellar mass BHs and IMBHs.  So, a detail study of the accretion process and the mass of accreting compact objects in ULXs will give more insight in understanding their nature and characteristics. 
	
The primary objective of this study is to gather recently acquired data from the {\it Chandra} ACIS-S detector, which involves extended exposure times, and analyze the spectral properties of non-nuclear X-ray point sources therein. This enables us to get a sample of 9 (nine) galaxies that were released in the public {\it Chandra} archive during the year 2019 to 2022. The distance to the galaxies were taken from available literature. The observation and data analysis are described in Section 2. Results and discussions are presented in Section 3 then summarized and concluded in Section 4.
	
\section{Observation and Data Analysis}\label{sec2}

For the present work, we have made use of $\it{Chandra}$ ACIS-S detector observed data. With the aim to work with relatively new observed data, we have considered suitable galaxies which have been observed during the year 2018 to 2021. This has resulted to the selection of an arbitrary 9 (nine) galaxies whose observation details are given in Table 1. The distance to the galaxies were obtained from available literature as listed in Table 1. The data reduction and analysis were done using Heasoft 6.29 $\&$ Chandra Interactive Analysis of Observations (CIAO) version 4.13. The X-ray point sources were extracted from the level 2 event lists by using the CIAO source detection tool {\it Wavdetect} which runs in two stages - {\it Wtransform} and {\it Wrecon}. Wtransform detects putative source pixels within a dataset through iterative correlation with "Mexican Hat" wavelet functions  with different wavlet scales of 1.0, 2.0, 4.0, 6.0, 8.0 and 16.0 pixels. Source pixels were identified by utilizing the default value of the parameter "sigthresh" set at $ \sim 10^{-6}$. {\it Wrecon} utilizes information obtained from {\it Wtransform} at each wavelet scale to generate a list of sources. The sources with net counts  $\gtrsim$100 were taken for the spectral analysis. The background regions were selected as a source free region near the corresponding sources.  A combination of CIAO tools and calibration data {\bf (CalDB 4.10.2)} were used to extract the source and background spectra. The spectra were grouped and binned at 15 counts per bin. Spectral Analysis were done by using the spectral fitting package XSPEC version 12.12.0. The spectra of all the sources were fitted in the 0.3-8.0 keV energy range by using two empirical models - (i) the multicolor disk blackbody model which represents emission from a standard Shakura \& Sunyaev (1973) \cite{shakura1973} thin disk and (ii) powerlaw model representing comptonization of soft photons near the accretion disc \cite{devi2007dependence,robba2021}, separately. Absorbtion was taken into account by using the XSPEC model phabs.  One of the best fitted source spectra (X-24; ObsID 23487), fitted with the two considered models are shown in Figure 1. While fitting the spectra, the hydrogen column density ($n_H$) was generally set free to vary, however for those cases when the estimated $n_H$ was much lower than the average Galactic value, it was frozen to the corresponding Galactic value \cite{HI4PI2016}. As many sources here have typically low number of counts, the analysis employed the C statistics. It is worth noting that, technically, C statistics are not recommended for high counts or background-subtracted data. However, a thorough examination revealed that the model parameters obtained using either C statistics or $\chi^2 $ statistics were consistently in agreement for sources with high count rates. This consistency is reassuring, as it suggests that the choice of statistical method did not significantly impact the results and provides added confidence in the results. Again, for the best fit, we compute the C-statistics for a range of parameters by using the STEPPAR command in XSPEC and thus the global minimum of C-statistics is found out.

\begin{table}[h]
\caption{Observation log of the sample Galaxies}\label{tab1}
\begin{tabular*}{\textwidth}{@{\extracolsep\fill}cccccc}
\toprule%
{ Galaxy }	&	{ ObsId }	&	{ Observation Date }	&	Exposure	&	Distance 	&	Distance Reference\\
&&(year-month-date)&(ks)&(Mpc)&\\
\midrule
{NGC 3079}	&	 19307	&	2018-01-30	&	53.16	&	15.6	&	\cite{devi2007dependence}\\
			&	20947	&	2018-02-01	&	44.40	&	&	\\
			\hline
			{PGC 032873}	&	21377	&	2019-01-22	&	57.31	&	108.8	&	\cite{buote2018luminous}\\
			&	22061	&	2019-01-26	&	31.65	&	&	\\
			&	21378	&	2019-02-11	&	24.74	&	&	\\
			&	22101	&	2019-02-13	&	33.62	&		&\\
			\hline
			{NGC 1316}	&	20340	&	2019-04-16	&	44.97	&	17	&	\cite{devi2007dependence}\\
			&	22179	&	2019-04-17	&	38.95	&	&	\\
			&	20341	&	2019-04-22	&	51.39	&	&	\\
			&	22187	&	2019-04-25	&	53.18	&	&	\\
			\hline
			
			{NGC 4472}	&	21647	&	2019-04-17	&	29.68	&	16.8	&	\cite{dage2019x}\\
			&	21648	&	2020-04-09	&	29.68	&	&	\\
			&	21649	&	2021-03-02	&	19.81	&	&	\\
			\hline
			
			IRAS 18293	&	21379	&	2019-08-08 	&	79.01	&	74.6		&	 \cite{randriamanakoto2013k}\\
			\hline
			{NGC 1600}	&	22878	&	2019-11-25	&	44.97	&	&	\cite{smith2008ngc}\\
			&	21375	&	2019-11-28	&	42.21	&	64.08	&	\\
			
			\hline
			NGC 7793	&	23266	&	2020-06-04 	&	29.69	&	3.90	&	\cite{israel2017discovery}\\
			\hline
			NGC 4214	&	22372	&	2020-08-01	&	59.28	&	2.90	&	\cite{fahrion2017disentangling}\\
			\hline
			
			{NGC 4485/90}	&	23482	&	2020-11-27 	&	26.68	&	7.8	&	\cite{devi2007dependence}\\
			&	23483	&	2020-12-27 	&	29.57	&	&	\\
			&	23484	&	2021-01-24 	&	29.57	&	&	\\
			&	23485	&	2021-02-21 	&	32.24	&	&	\\
			&	23486	&	2021-03-19 	&	29.66	&	&	\\
			&	23487	&	2021-04-18 	&	29.57	&	&	\\
			&	23490	&	2021-07-08 	&	29.57	&	&	\\
			&	23491	&	2021-08-06 	&	29.66	&	&	\\
\botrule

\end{tabular*}
\end{table}

\begin{figure}[h]
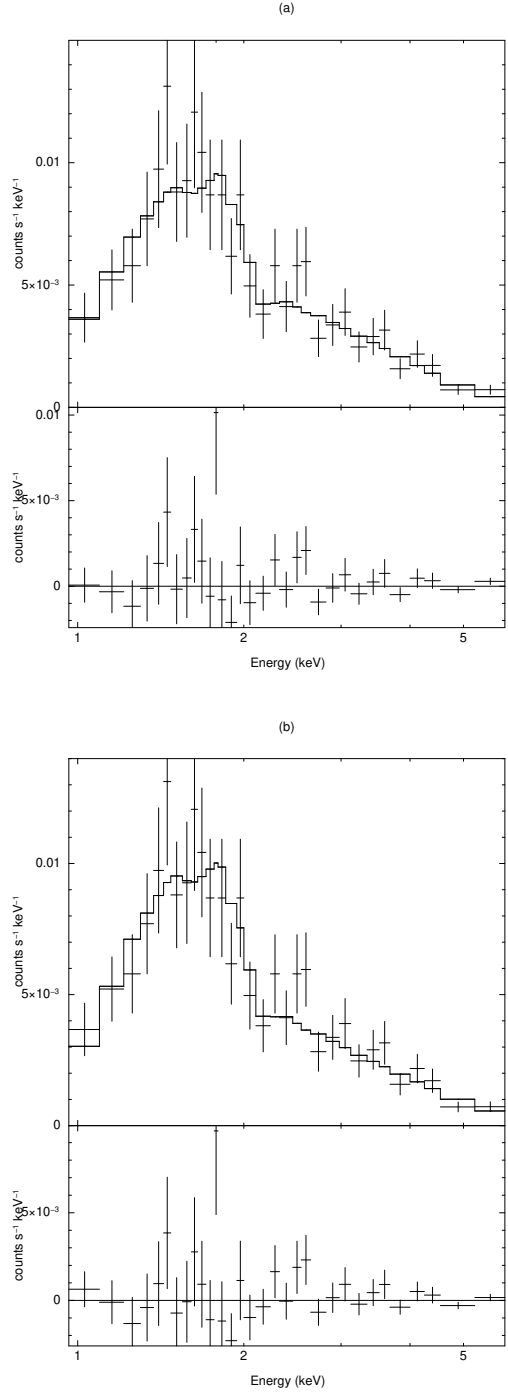
%
\centering
              \includegraphics*[width=0.55\textwidth,angle=0]{X-24-db-plot.eps} \\
              \includegraphics*[width=0.55\textwidth,angle=0]{X-24-pl-plot.eps} \\
\caption{Spectra of the source, X-24 (ObsId 23487) - (a) powerlaw modelled (b) disk blackbody modelled }
\end{figure}

\clearpage
\begin{table}[h]
 \fontsize{5}{10}\selectfont
\caption{Properties of the X-ray point sources}\label{tab2}
\begin{tabular*}{\textwidth}{@{\extracolsep\fill}ccccccc}
\toprule
{ Galaxy  }	&	{ Source }		&{ Source Name  }		&	{ R.A. }	&	Dec.	&{Uncetainty } & {Uncertainty}\\
&&(CXOU)&&&{in R.A.}&{in Dec.}\\
&&&&& {(seconds)} & {(arcseconds)}\\
\midrule
{NGC 3079}		&	X-1	&J100202.9+553859	&	10:02:02.87	&	+55:38:59.14	&0.004&0.032\\
			&	X-2	&	J100205.4+554258	&	10:02:05.36	&	+55:42:58.27	&0.002&0.026	\\
			&	X-3	&	J100201.0+553648	&10:02:00.95	&	+55:36:48.38	&0.008&0.060\\
			\hline	
			{PGC 032873}	&	X-4	&	J105604.8+421857 &	10:56:04.84	&	+42:18:57.34&0.002&0.035\\
			
			\hline
			{NGC 1316}	&	X-5	&	J032251.2-370950 &	03:22:51.24	&	-37:09:49.99& 0.006&0.082\\

			&	X-6	&		J032240.4-371640 &	03:22:40.43	&	-37:16:40.56&	0.004	&0.054\\
			
			&	X-7	&	J032240.8-371224 &		03:22:40.80	&	-37:12:23.66	&	0.003	&	0.038\\
			\hline
			
			NGC 4472	&	X-8	&	J122923.8+075359	&	12:29:23.77	&	+07:53:58.83	&0.004&0.052\\
			
			\hline
			
			{IRAS 18293-3413}	&	X-9	&	J183241.9-341625 &	18:32:41.89	&	-34:16:25.22	&0.004&0.081\\
			&	X-10	&	J183224.7-340917 &		18:32:24.66	&	-34:09:17.22&0.006&0.076\\	
			\hline
			
			{NGC 1600}	&	X-11	&	J043147.4-050213	&	04:31:47.43	&	-05:02:12.77	& 0.004&0.059\\
			
			&	X-12	&	J043139.7-050154	&	04:31:39.75	&	-05:01:54.38& 0.005&0.068	\\
			\hline

{NGC 7793 }	&	X-13&	J235806.7-323756 &		23:58:06.69	&	-32:37:56.44&0.003&0.032\\
			&	X-14	&	J235751.0-323726 &	23:57:50.99	&	-32:37:26.15	&0.016&0.020\\
			&	X-15	&	J235743.8-323634 &	23:57:43.83	&	-32:36:33.82	&0.003&0.050\\
			&	X-16	&	J235746.7-323607 &	23:57:46.75	&	-32:36:06.85&0.003&0.046\\
			\hline

			{NGC 4214}	&X-17	&	J121538.3+361920 &		12:15:38.27	&	+36:19:20.42&0.000&0.010\\
			&	X-18	&	J121538.2+361944 &		12:15:38.18	&	+36:19:44.13&0.002&0.040\\
			&	X-19	&		J121541.4+362114 &		12:15:41.42	&	+36:21:13.83&0.002&0.030\\
			\hline
			
			{NGC 4485/90}	&	X-20	&	J123043.1+413820 &		12:30:43.16	&	+41:38:19.65	&0.003&0.025\\
			
			&	X-21	&		J123036.2+413839 &	12:30:36.25	&	+41:38:38.92	&0.002&0.015\\
			
			&	X-22	&		J123035.1+413847 &	12:30:35.18	&	+41:38:47.49	&0.005&0.042\\
			&	X-23	&		J123030.7+413912 &	12:30:30.67	&	+41:39:12.38&0.001&0.016\\
			
			&	X-24	&	J123032.2+413919 &	12:30:32.20	&	+41:39:19.18	&0.001&0.014\\

			&	X-25	&	J123029.4+413928 &	12:30:29.42	&	+41:39:27.95	&0.003&0.038\\

			&	X-26	&	J123031.7+414142 &	12:30:31.66	&	+41:41:41.81		&0.002&0.021\\
			
			&	X-27	&	J123030.5+414142 &	12:30:30.48	&	+41:41:42.14	&0.003&0.038\\
\botrule
\end{tabular*}
\scriptsize {Notes:- Units of Right ascension (R.A.) are hours, minutes and seconds and that of Declination (Dec.) are degrees, arcminutes and arcseconds}

\end{table}
	
\section{Results and Discussions}\label{sec3}

In the present work, we have identified 27 X-ray point sources with net counts $\gtrsim$ 100 in the 9 selected galaxies. The spectral parameters of all the sources as estimated by the two models are tabulated in Table 3. For all the 27 sources the difference in (C statistics/degrees of freedom) of the two models were very small and hence we assume that both the models can equally explain the spectra of most of the sources except for few of them where one model is slightly preferred over the other. So, for the sources which are equally well explained by both the models, we consider the lower luminosity as estimated by the two models. With this criteria, we get 6 X-ray binaries ($L_X \sim 10^{38}$ erg s$^{-1}$) and 21 ULXs. Out of the 21 ULXs, we found 8 of them having even the lower limit of their luminosity greater than $10^{40}$ erg s$^{-1}$ and hence these 8 ULXs can be categorized as Extremely luminous X-ray sources (ELXs) \cite {devi2007dependence}. Figure 2 and Figure 3 respectively shows the variation of the luminosity with the inner disk temperature and powerlaw photon index. It is seen that  all the 6 X-ray binaries are in the spectrally hard state with their powerlaw photon index ($\Gamma$) ranging between 1.2 - 2.2. All the ULXs except for one ULX (X-5) were found to be in the spectrally hard state with their inner disk temperature kT$_{in} \gtrsim 1.0$ keV as estimated by the disk blackbody model and their powerlaw photon index ($\Gamma) \approx 2 $ within error limits. Only the ULX, X-5, is found to be spectrally soft with kT$_{in} \sim$ 0.4 keV and $\Gamma > 3.0 $. Thus from these plots, it can be inferred that neither the powerlaw photon index nor the inner disk temperature could significantly differentiate the X-ray binaries and the hard ULXs.

\begin{figure}[h!]%
\centering
\includegraphics[width=0.50\textwidth]{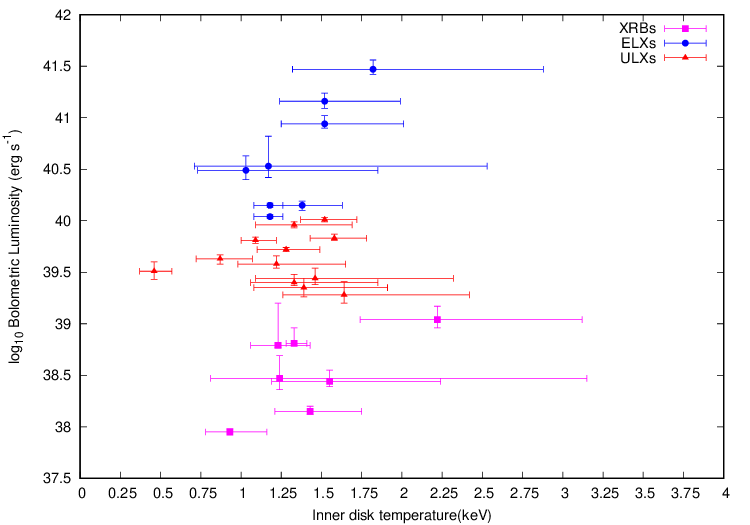}
\caption{Luminosity vs. Inner disk Temperature. (For sources having multiple observations, the best fitted ones have been considered in the plot.)}
\end{figure}

\begin{figure}[h]%
\centering
\includegraphics[width=0.50\textwidth]{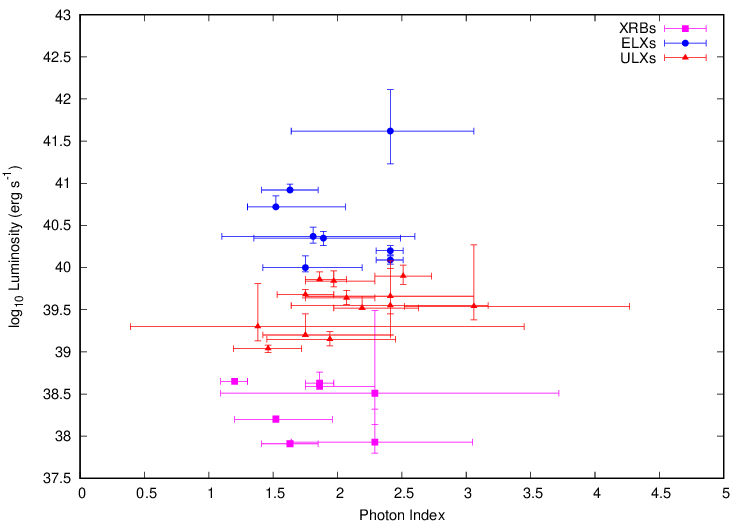}
\caption{Luminosity vs. Photon index. (For sources having multiple observations, the best fitted ones have been considered in the plot.)}
\end{figure}

\clearpage
\begin{figure}[h!]%
\centering
\includegraphics[width=1.0\textwidth]{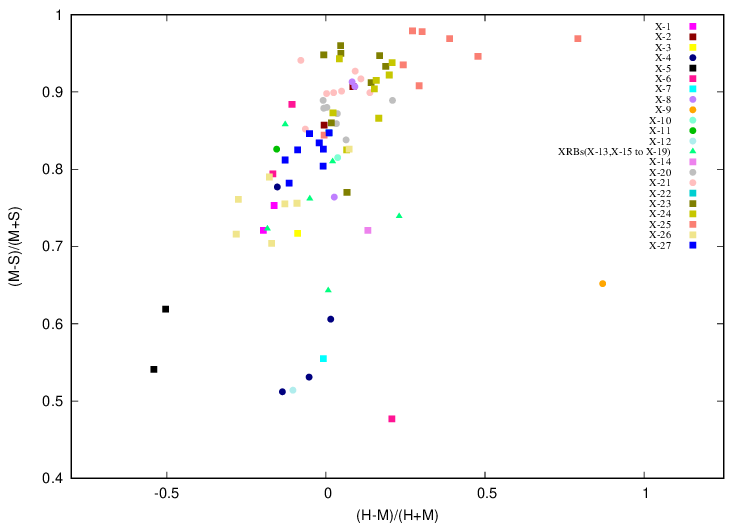}
\caption{Color-color plot of the X-ray point sources. (Square represents ULXs, triangle represents XRBs and circle represents ELXs.)}
\end{figure}
 As many of the sources in the present study are of low count, we tried to investigate their spectral properties by calculating the Hardness ratios. Background subtracted counts of all the sources were determined in the three energy bands: a soft band (S) of 0.3-1.0 keV, a medium band (M) of 1.0 - 2.0 keV and a hard band (H) of 2.0 - 8.0 keV. Then for each observation of every source, two types of hardness ratios were computed - (i) Soft color (SC) = $(M-S)/(M+S)$ and (ii) Hard color (HC) = $(H-M)/(H+M)$. Actually, the hardness ratio is not sufficient for definitively classifying individual sources. However, it can be utilized to explore general trends within a population of sources. Using the classification scheme of sources based on Hardness ratios as adopted by Kilgard et al. (2005) \cite{killgard}, we plotted the color-color diagram as shown in Figure 4. Consistent with the findings of \cite{killgard}, all 6 X-ray binaries in our study are observed to reside within the X-ray binary (XRB) hard state region, characterized by Hard color (HC) ranging between -0.2 to 0.3. Notably, in our investigation, these sources exhibit an even higher hardness, with Soft color (SC) ranging between 0.6 to 0.9. The ULXs are also found to have colors consistent with the XRBs. Among the 21 ULXs examined, 20 ULXs are identified to be situated in the hard spectral state region of the color-color diagram, characterized by HC $\sim$ 0.6 - 1.0 and SC $\sim$ 0.3 - 1.0 while only one ULX (X-5) was in a relatively soft spectral state with HC $\approx -0.5 $ and SC $\approx 0.5$. 
Even by examining the hardness ratios (HC/SC), it is evident that there is limited differentiation between the spectral states of hard X-ray binaries (XRBs) and hard ULXs/ELXs. Notably, two ULXs, namely X-20 and X-21, exhibit the peculiar behavior of manifesting the hardest state during their ELX regimes as compared to their ULX states across different observations.  \\
To investigate the overall characteristics of the 27 X-ray sources, the Cumulative luminosity functions (XLFs) of the sources are plotted in Figure 5. For this plot we have considered the lower limit of the bolometric luminosity for each source. Additionally, for sources with multiple observations, the best fitted one is taken. The XLF exhibits a faint break at around L$_X \sim 10^{40}$ erg s$^{-1}$, resulting in a limited number of X-ray sources (7 sources only) with minimum bolometric luminosity greater than $10^{40}$ erg s$^{-1}$. A similar break in the XLF around L$_X \sim 10^{40}$ erg s$^{-1}$ was also observed by Devi et al. (2007) \cite{devi2007dependence} in their study of around 365 X-ray sources using Chandra ACIS data. Additionally, the XLF reveals only two Hyperluminous X-ray sources (HLXs) with minimum L$_X > 10^{41}$ erg s$^{-1}$.

The galaxywise source properties are discussed in more details in the following sub-sections. 

\begin{figure}[h!]%
\centering
\includegraphics[width=0.7\textwidth]{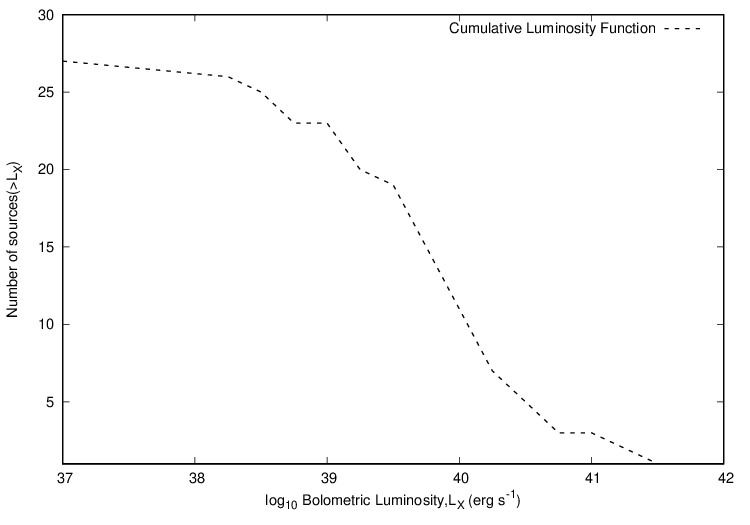}
\caption{Cumulative luminosity function of the X-ray point sources.}
\end{figure}

\clearpage
\setlength{\tabcolsep}{2pt}
\begin{sidewaystable*}
\centering
\caption{Spectral Properties of the X-ray sources}\label{tab3}
\begin{tabular*}{\textheight }{@{\extracolsep\fill}llc|cccc|cccr}
\toprule%
&&&\multicolumn{4}{@{}c@{}}{Disk Blackbody Model}&\multicolumn{4}{@{}c@{}}{Powerlaw Model} \\\cmidrule{4-7}\cmidrule{8-11}%
			Source  & Obs  &counts & $n_H$ & kT$_{in}$ & log(L$_X$) & c$_{stat}/d.o.f.$ &  $n_H$ & Gamma & log(L$_X$) & c$_{stat}/d.o.f.$\\
Number&ID&&(cm$^{-2}$)&(keV)&(erg s$^{-1}$)&&(cm$^{-2}$)&&(erg s$^{-1}$)&\\
\midrule
X-1	&	19307	&306	&$0.22^{+0.22}_{-0.20} \times 10^{22}$	&	$0.99^{+0.34}_{-0.20}$	&	$39.67^{+0.06}_{-0.05}$	&	6.84/14	&$0.54^{+0.31}_{-0.29}\times 10^{22}$	&$2.29^{+0.54}_{-0.54}$	&	$39.66^{+0.26}_{-0.15}$	&	7.61/14	\\
			&	20947	&	232	&$1.00^{+0.00}_{-0.99}\times 10^{20}$&	$1.22^{+0.43}_{-0.24}$	&	$39.58^{+0.08}_{-0.04}$	&	11.54/10	&$0.61^{+0.35}_{-0.41}\times 10^{22}$	&	$2.41^{+0.65}_{-0.66}$	&	$39.66^{+0.33}_{-0.21}$	&	9.84/10	\\
			
			X-2	&	19307	&	531	&$0.14^{+0.16}_{-0.12}\times 10^{20}$&	$1.33^{+0.36}_{-0.24}$	&	$39.96^{+0.03}_{-0.03}$	&	27.95/28	&$0.59^{+0.24}_{-0.21}\times 10^{22}$	&	$1.97^{+0.32}_{-0.22}$	&	$39.84^{+0.12}_{-0.07}$	&	19.66/28	\\
			&	20947	&	409	&$0.57^{+0.29}_{-0.28}\times 10^{22}$&	$1.28^{+0.27}_{-0.20}$	&	$40.01^{+0.05}_{-0.04}$	&	16.07/21	&$1.17^{+0.30}_{-0.43}\times 10^{22}$	&$2.29^{+0.44}_{-0.32}$	&	$40.04^{+0.22}_{-0.16}$	&	19.55/21 	\\
			
			X-3	&	19307	&	153	& *$8.74 \times 10^{19}$ &	$1.40^{+0.57}_{-0.34}$	&	$39.39^{+0.09}_{-0.09}$	&	5.10/5	&$0.07^{+0.53}_{-0.07}\times 10^{22}$	&$1.75^{+0.66}_{-0.33}$	&	$39.20^{+0.25}_{-0.02}$	&	5.72/5\\
			\midrule
			
			X-4	&21377	&	226	&	*$1.18\times 10^{20}$		&	$1.52^{+0.47}_{-0.28}$	&	$41.16^{+0.08}_{-0.07}$	&	12.14/10	&	*$1.18\times 10^{20}$	&	$1.58^{+0.24}_{-0.21}$	&	$40.91^{+0.04}_{-0.05}$	&	9.18/10	\\
			&22061	&		121	&	*$1.18\times 10^{20}$		&	$1.07^{+0.61}_{0.32}$	&	$41.12^{+0.11}_{-0.09}$	&	1.92/3	&	$0.09^{+10.32}_{0.09}\times 10^{20}$	&	$1.75^{+0.66}_{-0.44}$	&	$40.96^{+0.09}_{-0.00}$	&	1.38/3\\
	&21378	&		112	&	*$1.18\times 10^{20}$		&	$1.46^{+0.81}_{0.41}$	&	$41.22^{+0.13}_{-0.10}$	&	2.12/3	&	$0.01^{+0.46}_{0.01}\times 10^{22}$	&	$1.63^{+0.65}_{0.33}$	&	$40.99^{+0.20}_{-0.00}$	&	0.59/3		\\
&	22101	&131	&	*$1.18\times 10^{20}$	&	$1.41^{+0.89}_{0.38}$	&	$41.16^{+0.15}_{-0.10}$	&	1.17/4 	&	$0.28^{+0.61}_{0.28}\times 10^{22}$	&	$1.86^{+0.88}_{-0.65}$	&	$41.02^{+0.34}_{-0.09}$	&	0.68 /4	\\
		\midrule

			X-5&20341		&	179	&	*$1.90 \times 10^{20}$	&	$0.46^{+0.11}_{-0.09}$	&	$39.51^{+0.09}_{-0.08}$	&	4.94/6	&	$0.05^{+0.41}_{-0.05} \times 10^{20}$	&	$3.06^{+1.21}_{-0.54}$	&	$39.54^{+0.73}_{-0.16}$	&	2.15/6	\\
			&22187			&	147	&	*$1.90 \times 10^{20}$	&	$0.41^{+0.22}_{-0.05}$	&	$39.46^{+0.11}_{-0.14}$	&	11.28/4	&	$0.10^{+9.90}_{-0.09}\times 10^{20}$	&	$3.04^{+0.77}_{-0.78}$	&	$39.44^{+0.28}_{-0.19}$	&	6.54/4	\\	
			X-6	&22179	&		191	&	*$1.90 \times 10^{20}$	&	$1.23^{+0.33}_{-0.26}$ 	&	$39.72^{+0.06}_{-0.07}$	&	8.82/8	&	$0.03^{+0.07}_{-0.03}\times 10^{22}$	&	$1.97^{+0.32}_{-0.33}$ 	&	$39.56^{+0.07}_{-0.02}$	&	7.20/8	\\		
			&20341	&		179	&	*$1.90 \times 10^{20}$	&	$0.87^{+0.20}_{-0.15}$	&	$39.63^{+0.04}_{-0.05}$	&	13.38/10	&	$0.09^{+14.01}_{-0.09}\times 10^{22}$	&	$2.19^{+0.44}_{-0.22}$	&	$39.52^{+0.14}_{-0.02}$	&	7.34/10	\\		
			
			&22187	&	207	&	*$1.90 \times 10^{20}$	&	$1.78^{+1.74}_{-0.53}$	&	$39.57^{+0.26}_{-0.11}$	&	29.94/8	&	$0.01^{+0.00}_{-0.00}\times 10^{22}$	&	$1.47^{+0.42}_{-0.37}$	&	$39.31^{+0.05}_{-0.07}$	&	25.88/8	\\		
			X-7	&22187	&	160	&	*$1.90\times 10^{20}$	&	$1.39^{+0.52}_{-0.31}$	&	$39.35^{+0.09}_{-0.09}$	&	3.68/5	&	$0.49^{+0.89}_{-0.49}\times 10^{22}$	&	$1.38^{+2.07}_{-0.99}$	&	$39.30^{+0.51}_{-0.17}$	&	4.64/5	\\		
			\midrule
			X-8&	21647	&	257	&	*$1.55\times 10^{20}$	&	$1.89^{1.03}_{-0.45}$	&	$40.07^{+0.15}_{-0.09}$	&	4.85/12	&$0.10^{+0.81}_{-0.10}\times 10^{22}$	&	$1.41^{+0.77}_{-0.32}$	&	$39.80^{+0.24}_{-0.02}$	&	5.27/12\\
			&	21648	&	365	&		*$1.55 \times 10^{20}$	&	$1.38^{+0.25}_{-0.21}$	&	$40.15^{+0.04}_{-0.05}$	&	18.92/18	&	$0.18^{+0.33}_{-0.18} \times 10^{22}$	&$1.75^{+0.44}_{-0.33}$	&	$40.00^{+0.14}_{-0.05}$	&	15.03/18\\
			&	21649	&	292	&		*$1.55 \times 10^{20}$	&		$1.53^{+0.43}_{-0.26}$	&	$40.24^{+0.06}_{-0.06}$	&	24.06/14	&	$0.10^{+9.9}_{-0.09} \times 10^{20}$	&	$1.75^{+0.11}_{-0.22}$	&	$40.05^{+0.03}_{-0.01}$	&	18.89/14\\
	\midrule		
X-9	&	21379	&	305	&	$4.04^{+1.53}_{-1.37} \times 10^{22}$	&	$1.82^{+1.06}_{-0.50}$	&	$41.47^{+0.09}_{-0.05}$	&	11.17/15	&	$5.92^{+1.53}_{-1.72} \times 10^{22}$	&	$2.41^{+0.65}_{-0.77}$	&	$41.62^{+0.49}_{-0.39}$	&	10.18/15	\\
X-10	&&		300	&	*$1.07 \times 10^{20}$	&	$1.52^{+0.49}_{-0.27}$	&	$40.94^{+0.08}_{-0.04}$	&	13.93/13	&	$0.05^{+0.40}_{-0.05} \times 10^{20}$	&	$1.52^{+0.54}_{-0.22}$	&	$40.72^{+0.13}_{-0.01}$	&	12.42/13	\\

\botrule	
\end{tabular*}
{\scriptsize Note:-  * hydrogen column density frozen to the average galactic value }
\end{sidewaystable*}	

\setlength{\tabcolsep}{2pt}
\begin{sidewaystable*}
\centering
\begin{tabular*}{\textheight }{@{\extracolsep\fill}llc|cccc|cccr}
\toprule%
&&&\multicolumn{4}{@{}c@{}}{Disk Blackbody Model}&\multicolumn{4}{@{}c@{}}{Powerlaw Model} \\\cmidrule{4-7}\cmidrule{7-11}%
			Source  & Obs  &counts & $n_H$ & kT$_{in}$ & log(L$_X$) & c$_{stat}/d.o.f.$ & $n_H$ & Gamma & log(L$_X$) & c$_{stat}/d.o.f.$\\
Number&ID&&(cm$^{-2}$)&(keV)&(erg s$^{-1}$)&&(cm$^{-2}$)&&(erg s$^{-1}$)&\\
\midrule
X-11	&	22878	&	138	&	*$3.14 \times 10^{20}$	&	$1.17^{+1.36}_{-0.46}$	&	$40.53^{+0.29}_{-0.11}$	&	8.47/3	&	*$3.14 \times 10^{20}$	&	$1.81^{+0.79}_{-0.71}$	&	$40.37^{+0.11}_{-0.08}$	&	6.12/3\\
	X-12	&	21375	&	123	&	*$3.14 \times 10^{20}$	&	$1.03^{+0.82}_{-0.30}$	&	$40.49^{+0.14}_{-0.09}$	&	5.33/3 	&*$3.14 \times 10^{20}$	&$1.89^{+0.60}_{-0.54}$	&	$40.35^{+0.08}_{-0.09}$	&	2.54/3\\

			\midrule
		X-13 	&23266	&	333	&	$0.10^{+0.90}_{-0.09} \times 10^{20}$	&	$1.23^{+0.20}_{-0.17}$	&	$38.79^{+0.41}_{-0.02}$	&	21.66/16	&	$0.00^{+0.37}_{-0.00}\times 10^{22}$	&	$1.86^{+0.43}_{-0.11}$	&	$38.59^{+0.17}_{-0.00}$	&	12.85/16	\\
			
			X-14	&&	403	&	*$3.32\times 10^{20}$	&	$2.22^{+0.90}_{-0.48}$	&	$39.04^{+0.13}_{-0.08}$	&	30.07/21	&	$0.02^{+0.07}_{-0.02}\times 10^{22}$	&	$1.20^{+0.10}_{-0.11}$	&	$38.65^{+0.02}_{-0.01}$	&	26.95/21	\\
			
X-15	&&	116	&	$0.90^{+1.16}_{-0.82}\times 10^{22}$	&	$1.24^{+1.91}_{-0.43}$	&	$38.47^{+0.22}_{-0.11}$	&	13.63/3	&	$1.55^{+1.54}_{-1.53}\times 10^{22}$	&	$2.29^{+1.43}_{-1.20}$	&	$38.51^{+0.98}_{-0.37}$	&	14.14/3	\\
			X-16	&&148	&	$0.09^{+0.99}_{-0.08}\times 10^{20}$	&	$1.55^{+0.69}_{-0.36}$	&	$38.44^{+0.11}_{-0.05}$	&	3.99/5	&	$0.10^{+10.81}_{-0.10}\times 10^{20}$	&	$1.52^{+0.44}_{-0.32}$	&	$38.20^{+0.04}_{-0.00}$	&	3.41/5	\\
			
		\midrule	
			X-17	&22372	&	1347	&	$0.10^{+0.90}_{0.09}\times 10^{20}$	&	$1.33^{+0.08}_{-0.05}$	&	$38.81^{+0.15}_{-0.00}$	&	86.11/70	&	$0.15^{+0.09}_{-0.08}\times 10^{22}$	&	$1.86^{+0.11}_{-0.11}$	&	$38.63^{+0.03}_{-0.03}$	&	80.85/70	\\
			X-18	&&		216	&	*$3.88\times 10^{20}$	&		$0.94^{+0.24}_{-0.17}$	&	$37.94^{+0.05}_{-0.06}$	&	7.55/9	&	$0.26^{+0.47}_{-0.26}\times 10^{22}$	&	$2.29^{+0.76}_{-0.65}$	&	$37.93^{+0.39}_{-0.13}$	&	7.77/9	\\
			X-19	&		&	299	&$0.10^{+0.90}_{-0.01}\times 10^{20}$	&	$1.43^{+0.32}_{-0.22}$	&	$38.15^{+0.05}_{-0.03}$	&	26.57/14	&	$0.24^{+10.66}_{-0.24}\times 10^{20}$	&	$1.63^{+0.22}_{-0.22}$	&	$37.91^{+0.03}_{-0.00}$	&	20.32/14	\\
\midrule
	X-20	&23482	&	549	&	*$3.78\times 10^{20}$	&	$1.69^{+0.26}_{-0.22}$	&	$39.67^{+0.05}_{-0.04}$	&	26.61/29	&	$0.42^{+0.24}_{-0.18}\times 10^{22}$	&	$1.86^{+0.33}_{-0.22}$	&	$39.54^{+0.11}_{-0.06}$	&	25.84/29	\\
			&23483	&	548	&	*$3.78\times 10^{20}$	&	$1.76^{+0.35}_{-0.23}$	&	$39.68^{+0.05}_{-0.04}$	&	29.60/29	&	$0.42^{+0.24}_{-0.23}\times 10^{22}$	&	$1.86^{+0.21}_{-0.22}$	&	$39.53^{+0.09}_{-0.07}$	&	22.95/29	\\
			&23484	&715	&	$0.30^{+0.17}_{-0.16}\times 10^{22}$	&	$1.18^{+0.16}_{-0.12}$	&	$39.80^{+0.03}_{-0.03}$	&	26.83/37	&	$0.90^{+0.23}_{-0.18}\times 10^{22}$	&	$2.51^{+0.22}_{-0.22}$	&	$39.90^{+0.13}_{-0.11}$	&	22.44/37	\\
			&23485	&1106	&	$0.15^{+0.11}_{-0.11}\times 10^{22}$	&	$1.29^{+0.14}_{-0.10}$	&	$39.95^{+0.02}_{-0.02}$	&	54.26/55	&	$0.65^{+0.16}_{-0.13}\times 10^{22}$	&	$2.29^{+0.22}_{-0.10}$	&	$39.95^{+0.10}_{-0.06}$	&	42.66/55	\\
			&23486	&	1276	&	$0.30^{+0.09}_{-0.10}\times 10^{22}$	&	$1.18^{+0.08}_{-0.10}$	&	$40.04^{+0.02}_{-0.02}$	&	54.35/65	&	$0.80^{+0.09}_{-0.08}\times 10^{22}$	&$2.41^{+0.10}_{-0.11}$	&	$40.09^{+0.05}_{-0.05}$	&	59.99/65	\\
			&23487	&	757	&	$4.97^{+14.83}_{-4.40}\times 10^{20}$	&	$1.39^{+0.19}_{-0.16}$	&	$39.78^{+0.02}_{-0.02}$	&	47.47/42	&	$0.50^{+0.20}_{-0.18}\times 10^{22}$	&	$2.19^{+0.22}_{-0.22}$	&	$39.74^{+0.11}_{-0.08}$	&	42.33/42	\\
			&23491	&	834		&	$0.27^{+0.13}_{-0.16}\times 10^{22}$	&	$1.52^{+0.16}_{-0.12}$	&	$39.84^{+0.03}_{-0.03}$	&	40.38/44	&	$0.70^{+0.18}_{-0.20}\times 10^{22}$	&	$2.29^{+0.22}_{-0.22}$	&	$39.85^{+0.11}_{-0.09}$	&	49.27/44	\\
\midrule
X-21	&	23482		&448	&	$0.66^{+0.44}_{-0.38}\times 10^{22}$	&	$0.76^{+0.18}_{-0.13}$	&	$39.69^{+0.15}_{-0.11}$	&	14.29/23	&	$1.55^{+0.64}_{-0.57}\times 10^{22}$	&	$3.50^{+0.66}_{-0.66}$	&	$40.31^{+0.54}_{-0.46}$	&	14.55/23	\\
			&23483	&580	&	$0.31^{+0.18}_{-0.18}\times 10^{22}$	&	$0.96^{+0.14}_{-0.10}$	&	$39.69^{+0.05}_{-0.04}$	&	21.63/30	&	$0.82^{+0.21}_{-0.24}\times 10^{22}$	&	$2.63^{+0.43}_{-0.33}$	&	$39.82^{+0.23}_{-0.16}$	&	29.17/30		\\
			&23484	&856	&	$0.39^{+0.14}_{-0.15}\times 10^{22}$	&	$1.00^{+0.12}_{-0.08}$	&	$39.90^{+0.03}_{-0.03}$	&	28.45/45	&	$0.93^{+0.11}_{-0.19}\times 10^{22}$	&	$2.63^{+0.21}_{-0.22}$	&	$40.04^{+0.11}_{-0.12}$	&	40.71/45	\\
			&23485	&	1561	&	$0.51^{+0.13}_{-0.12}\times 10^{22}$	&	$1.18^{+0.08}_{-0.10}$	&	$40.15^{+0.02}_{-0.02}$	&	81.11/80	&	$1.05^{+0.12}_{-0.11}\times 10^{22}$	&	$2.41^{+0.10}_{-0.11}$	&	$40.20^{+0.06}_{-0.05}$	&	102.01/80	\\
			&23486	&		1088	&	$0.20^{+0.12}_{-0.12}\times 10^{22}$	&	$1.41^{+0.15}_{-0.11}$	&	$39.97^{0.02}_{-0.02}$	&	62.75/57	&	$0.66^{+0.17}_{-0.13}\times 10^{22}$	&	$2.07^{+0.22}_{-0.10}$	&	$39.90^{+0.09}_{-0.05}$	&	60.46/57	\\
			&23487	&		1448	&	$0.30^{+0.13}_{-0.10}\times 10^{22}$	&	$1.39^{+0.12}_{-0.14}$	&	$40.13^{+0.01}_{-0.01}$	&	85.20/75	&	$0.75^{+0.09}_{-0.15}\times 10^{22}$	&	$2.07^{+0.11}_{-0.10}$	&	$40.06^{+0.04}_{-0.05}$	&	95.83/75	\\
			&23490	&	912	&	$0.32^{+0.14}_{-0.15}\times 10^{22}$	&	$1.24^{+0.14}_{-0.10}$	&	$39.92^{+0.02}_{-0.02}$	&	29.41/48	&	$0.87^{+0.22}_{-0.17}\times 10^{22}$	&	$2.41^{+0.22}_{-0.22}$	&	$39.97^{+0.12}_{-0.10}$	&	38.84/48	\\
			&23491	&		962	&	$0.37^{+0.11}_{-0.14}\times 10^{22}$	&	$1.05^{+0.12}_{-0.06}$	&	$39.93^{+0.02}_{-0.03}$	&	38.38/49	&	$0.85^{+0.10}_{-0.17}\times 10^{22}$	&	$2.51^{+0.22}_{-0.22}$	&	$40.01^{+0.10}_{-0.10}$	&	48.55/49	\\

\botrule	
\end{tabular*}
{\scriptsize Note:-  * hydrogen column density frozen to the average galactic value }
\end{sidewaystable*}
\setlength{\tabcolsep}{2pt}
\begin{sidewaystable*}
\centering
\begin{tabular*}{\textheight }{@{\extracolsep\fill}llc|cccc|cccr}
\toprule%
&&&\multicolumn{4}{@{}c@{}}{Disk Blackbody Model}&\multicolumn{4}{@{}c@{}}{Powerlaw Model} \\\cmidrule{4-7}\cmidrule{8-11}%
			Source  & Obs  &counts & $n_H$ & kT$_{in}$ & log(L$_X$) & c$_{stat}/d.o.f.$ & $n_H$ & Gamma & log(L$_X$) & c$_{stat}/d.o.f.$\\
Number&ID&&(cm$^{-2}$)&(keV)&(erg s$^{-1}$)&&(cm$^{-2}$)&&(erg s$^{-1}$)&\\
\midrule

			X-22	&23484	&238	&	*$3.78 \times 10^{20}$	&	$1.64^{+0.78}_{-0.38}$	&	$39.28^{+0.13}_{-0.08}$	&	8.97/10 	&	*$3.78 \times 10^{20}$	&	$1.46^{+0.26}_{-0.27}$	&	$39.04^{+0.04}_{-0.05}$	&	5.08/10 \\
			\midrule
			X-23	&23482	&	1093	&	$0.43^{+0.13}_{-0.13}\times 10^{22}$	&	$1.08^{+0.10}_{-0.09}$	&	$40.02^{+0.03}_{-0.03}$	&	45.41/58	&	$1.12^{+0.13}_{-0.12}\times 10^{22}$	&	$2.74^{+0.22}_{-0.10}$	&	$40.23^{+0.12}_{-0.07}$	&	42.69/58	\\
			&	23483	&935	&	*$3.78 \times 10^{20}$	&	$1.58^{+0.17}_{-0.12}$	&	$39.90^{+0.03}_{-0.03}$	&	39.94/49		&	$0.93^{+0.24}_{-0.27}\times 10^{22}$	&	$2.29^{+0.33}_{-0.22}$	&	$39.97^{+0.16}_{-0.11}$	&	29.95/49	\\
			&23484	&1065	&	$0.38^{+0.14}_{-0.15}\times 10^{22}$	&	$1.19^{+0.13}_{-0.10}$	&	$39.99^{+0.02}_{-0.02}$	&	48.93/55	&	$1.04^{+0.12}_{-0.21}\times 10^{22}$	&	$2.51^{+0.22}_{-0.22}$	&	$40.11^{+0.11}_{-0.11}$	&	49.84/55	\\
			&23485	&184	&	*$3.78 \times 10^{20}$	&	$1.58^{+0.71}_{-0.37}$	&	$39.17^{+0.11}_{-0.08}$	&	2.96/7		&	$0.04^{+0.11}_{-0.04}\times 10^{22}$	&	$1.52^{+0.33}_{-0.32}$	&	$38.94^{+0.03}_{-0.01}$	&	2.17/7	\\
			&23486	&	918	&	$0.28^{+0.16}_{-0.17}\times 10^{22}$	&	$1.64^{+0.29}_{-0.19}$	&	$39.95^{+0.02}_{-0.02}$	&	49.35/50	&	$0.80^{+0.20}_{-0.23}\times 10^{22}$	&	$1.97^{+0.22}_{-0.22}$	&	$39.84^{+0.09}_{-0.07}$	&	47.43/50	\\
			&23487	&896	&	$0.44^{+0.14}_{-0.15}\times 10^{22}$	&	$1.41^{+0.19}_{-0.14}$	&	$39.94^{+0.02}_{-0.02}$	&	59.65/49	&	$1.05^{+0.12}_{-0.21}\times 10^{22}$	&	$2.29^{+0.11}_{-0.22}$	&	$39.95^{+0.06}_{-0.10}$	&	52.00/49	\\
			&23490	&773	&	$0.27^{+0.14}_{-0.14}\times 10^{22}$	&	$1.09^{+0.13}_{-0.09}$	&	$39.81^{+0.03}_{-0.03}$	&	37.18/41	&	$0.80^{+0.20}_{-0.16}\times 10^{22}$	&	$2.51^{+0.22}_{-0.22}$	&	$39.90^{+0.13}_{-0.10}$	&	41.41/41	\\
			&23491	&	674	&	$0.52^{+0.23}_{-0.24}\times 10^{22}$	&	$1.40^{+0.26}_{-0.19}$	&	$39.83^{+0.03}_{-0.03}$	&	43.10/35	&	$1.17^{+0.30}_{-0.34}\times 10^{22}$	&	$2.29^{+0.33}_{-0.32}$	&	$39.86^{+0.17}_{-0.14}$	&	40.45/35	\\
\midrule
X-24	&23482	&	1083	&	$0.35^{+0.13}_{-0.13}\times 10^{22}$	&	$1.58^{+0.20}_{-0.15}$	&	$40.05^{+0.02}_{-0.02}$	&	68.90/59	&$0.82^{+0.21}_{-0.16}\times 10^{22}$		&	$1.97^{+0.22}_{-0.11}$	&	$39.93^{+0.09}_{-0.05}$	&	70.51/59	\\
			&23483	&1028	&	$0.40^{+0.15}_{-0.15}\times 10^{22}$	&	$1.40^{+0.15}_{-0.14}$	&	$40.00^{+0.02}_{-0.02}$	&	44.95/53	&	$0.91^{+0.11}_{-0.18}\times 10^{22}$		&	$2.19^{+0.10}_{-0.22}$	&	$39.97^{+0.05}_{-0.08}$	&	49.78/53	\\
			&23484	&	1013	&	$0.45^{+0.14}_{-0.15}\times 10^{22}$	&	$1.35^{+0.18}_{-0.13}$	&	$39.99^{+0.02}_{-0.02}$	&	53.11/55	&	$0.90^{+0.23}_{-0.18}\times 10^{22}$		&	$2.07^{+0.22}_{-0.10}$	&	$39.92^{+0.10}_{-0.06}$	&	65.21/55	\\
			&23485	&	1031	&	$0.46^{+0.14}_{-0.00}\times 10^{22}$	&	$1.19^{+0.13}_{-0.10}$	&	$40.00^{+0.02}_{-0.02}$	&	61.41/54	&	$1.07^{+0.13}_{-0.22}\times 10^{22}$		&	$2.51^{+0.11}_{-0.22}$	&	$40.10^{+0.07}_{-0.11}$	&	69.12/54	\\
			&23486	&	1082&	$0.33^{+0.12}_{-0.12}\times 10^{22}$	&	$1.52^{+0.20}_{-0.15}$	&	$40.01^{+0.02}_{-0.02}$	&	53.73/59	&	$0.62^{+0.25}_{-0.06}\times 10^{22}$		&	$1.86^{+0.21}_{-0.11}$	&	$39.86^{+0.09}_{-0.02}$	&	64.05/59	\\
			&23487	&	550	&	$0.25^{+0.21}_{-0.20}\times 10^{22}$	&	$1.35^{+0.28}_{-0.18}$	&	$39.67^{+0.03}_{-0.03}$	&	29.55/29	&	$0.75^{+0.31}_{-0.27}\times 10^{22}$		&	$2.19^{+0.32}_{-0.33}$	&	$39.64^{+0.16}_{-0.12}$	&	28.66/29		\\
			&23490	&	741	&	*$3.78\times 10^{20}$		&	$2.09^{+0.34}_{-0.22}$	&	$39.84^{+0.05}_{-0.04}$	&	65.04/40	&	$1.31^{+0.34}_{-0.27}\times 10^{22}$		&	$2.29^{+0.33}_{-0.22}$	&	$39.93^{+0.18}_{-0.11}$	&	49.41/40	\\
			&23491	&	476	&	$0.10^{+0.22}_{-0.09}\times 10^{22}$	&	$1.52^{+0.35}_{-0.23}$	&	$39.62^{+0.03}_{-0.03}$	&	15.13/25	&	$0.64^{+0.26}_{-0.38}\times 10^{22}$	&	$2.07^{+0.22}_{-0.43}$	&	$39.55^{+0.11}_{-0.13}$	&	13.33/25	\\
\midrule
X-25	&23482	&	268	&	$1.07^{+0.79}_{-0.85}\times 10^{22}$	&	$1.13^{+0.68}_{-0.28}$	&	$39.56^{+0.15}_{-0.09}$	&	19.92/12	&	$2.01^{+1.18}_{-1.21}\times 10^{22}$	&	$2.63^{+0.98}_{-0.88}$	&	$39.78^{+0.72}_{-0.42}$	&	20.29/12		\\
			&23483	&	289	&$2.10^{+0.80}_{-0.71}\times 10^{22}$	&	$1.00^{+0.31}_{-0.18}$	&	$39.74^{+0.14}_{-0.11}$	&	16.46/14	&	$3.42^{+0.88}_{-1.00}\times 10^{22}$	&	$3.18^{+0.76}_{-0.76}$	&	$40.29^{+0.63}_{-0.52}$	&	17.29/14	\\
			&23484	&	188	&	$2.01^{+1.17}_{-1.09}\times 10^{22}$	&	$0.86^{+0.42}_{-0.21}$	&	$39.60^{+0.28}_{-0.19}$	&	4.80/7	&	$3.39^{+1.40}_{-1.48}\times 10^{22}$	&	$3.61^{+1.21}_{-1.20}$	&	$40.40^{+1.06}_{-0.85}$	&	5.74/7	\\
			&23485	&	271	&	$0.91^{+0.67}_{-0.66}\times 10^{22}$	&	$1.28^{+0.45}_{-0.27}$	&	$39.47^{+0.09}_{-0.07}$	&	24.80/13	&	$2.03^{+0.83}_{-0.89}\times 10^{22}$	&	$2.74^{+0.66}_{-0.66}$	&	$39.76^{+0.49}_{-0.37}$	&	24.29/13	\\
			&23486	&248	&	*$3.78\times 10^{20}$		&	$1.54^{+0.47}_{-0.26}$	&	$39.35^{+0.07}_{-0.06}$	&	5.95/11		&	$1.20^{+0.94}_{-1.08}\times 10^{22}$	&	$2.29^{+0.76}_{-0.76}$	&	$39.45^{+0.48}_{-0.30}$	&	5.48/11	\\
			&23487	&223	&	$0.84^{+0.84}_{-0.76}\times 10^{22}$	&	$1.46^{+0.86}_{-0.37}$	&	$39.44^{+0.10}_{-0.06}$	&	10.23/10	&	$1.89^{+1.10}_{-1.22}\times 10^{22}$	&	$2.41^{+0.76}_{-0.77}$	&	$39.55^{+0.53}_{-0.35}$	&	10.68/10	\\
			&23490	&478	&	*$3.78\times 10^{20}$		&	$1.45^{+0.26}_{-0.20}$	&	$39.52^{+0.05}_{-0.04}$	&	41.50/25	&	$0.05^{+0.24}_{-0.05}\times 10^{22}$	&	$1.63^{+0.33}_{-0.22}$	&	$39.34^{+0.08}_{-0.01}$	&	38.03/25		\\
			&23491	&243	&	$1.14^{+0.75}_{-0.68}\times 10^{22}$	&	$1.20^{+0.54}_{-0.28}$	&	$39.56^{+0.12}_{-0.07}$	&	10.76/11	&	$2.24^{+0.92}_{-0.98}\times 10^{22}$	&	$2.74^{+0.76}_{-0.76}$	&	$39.85^{+0.57}_{-0.41}$	&	9.66/11	\\
\botrule	
\end{tabular*}
{\scriptsize Note:-  * hydrogen column density frozen to the average galactic value }
\end{sidewaystable*}

\setlength{\tabcolsep}{2pt}
\begin{sidewaystable*}
\centering
\begin{tabular*}{\textheight }{@{\extracolsep\fill}llc|cccc|cccr}
\toprule%
&&&\multicolumn{4}{@{}c@{}}{Disk Blackbody Model}&\multicolumn{4}{@{}c@{}}{Powerlaw Model} \\\cmidrule{4-7}\cmidrule{8-11}%
			Source  & Obs  &counts & $n_H$ & kT$_{in}$ & log(L$_X$) & c$_{stat}/d.o.f.$ &  $n_H$ & Gamma & log(L$_X$) & c$_{stat}/d.o.f.$\\
Number&ID&&(cm$^{-2}$)&(keV)&(erg s$^{-1}$)&&(cm$^{-2}$)&&(erg s$^{-1}$)&\\

\midrule

X-26	&23482	&	243	&	*$3.78\times 10^{20}$		&	$1.14^{+0.25}_{-0.18}$	&	$39.22^{+0.05}_{-0.04}$	&	9.81/11	&	$0.14^{+0.42}_{-0.14}\times 10^{22}$	&	$1.97^{0.54}_{-0.44}$	&	$39.11^{+0.23}_{-0.04}$	&	11.34/11	\\
			&23483	&	224	&	*$3.78\times 10^{20}$		&	$0.82^{+0.19}_{-0.14}$	&	$39.11^{+0.05}_{-0.05}$	&	16.76/10	&	$0.63^{+10.73}_{-0.63}\times 10^{20}$	&	$2.19^{+0.44}_{-0.33}$	&	$39.02^{+0.12}_{-0.02}$	&	16.52/10	\\
			&23484	&	257	&	*$3.78\times 10^{20}$		&	$0.91^{+0.16}_{-0.12}$	&	$39.21^{+0.04}_{-0.05}$	&	20.38/12	&	$0.32^{+0.32}_{-0.32}\times 10^{22}$	&	$2.63^{+0.43}_{-0.55}$	&	$39.31^{+0.27}_{-0.23}$	&	22.99/12\\
			&	23485	&	320	&	*$3.78\times 10^{20}$		&	$0.87^{+0.15}_{-0.11}$	&	$39.24^{+0.04}_{-0.04}$	&	26.46/16	&	*$3.78\times 10^{20}$	&	$2.16^{+0.26}_{-0.23}$	&	$39.12^{+0.05}_{-0.05}$	&	22.51/16	\\
			&23486	&	893	&	*$3.78\times 10^{20}$	&	$1.58^{+0.20}_{-0.15}$	&	$39.83^{+0.04}_{-0.02}$	&	50.63/49	&	$0.30^{+0.17}_{-0.16}\times 10^{22}$	&	$1.75^{+0.22}_{-0.22}$	&	$39.68^{+0.06}_{-0.04}$	&	47.78/49	\\
			&23487	&	301	&	*$3.78\times 10^{20}$		&	$1.04^{+0.18}_{-0.13}$	&	$39.36^{+0.04}_{-0.04}$	&	8.15/14		&	$0.38^{+0.38}_{-0.32}\times 10^{22}$	&	$2.29^{+0.54}_{-0.43}$	&	$39.37^{+0.28}_{-0.15}$	&	9.54/14	\\
			&23490	&	213	&	*$3.78\times 10^{20}$		&	$1.09^{+0.34}_{-0.22}$	&	$39.16^{+0.05}_{-0.06}$	&	9.56/8	&	$0.10^{+9.90}_{-0.00}\times 10^{20}$	&	$1.75^{+0.44}_{-0.22}$	&	$38.99^{+0.06}_{-0.00}$	&	8.82/8	\\
			\midrule
			X-27	&23482	&341	&	*$3.78\times 10^{20}$		&	$1.13^{+0.13}_{-0.14}$	&	$39.36^{+0.04}_{-0.03}$	&	15.15/17	&	$0.34^{+0.26}_{-0.30}\times 10^{22}$	&	$2.29^{+0.33}_{-0.32}$	&	$39.35^{+0.17}_{-0.14}$	&	21.45/17	\\
			&23483	&	691	&	$8.45^{+14.83}_{-7.57}\times 10^{20}$	&	$1.28^{+0.21}_{-0.18}$	&	$39.72^{+0.02}_{-0.02}$	&	38.28/37	&	$0.41^{+0.17}_{-0.18}\times 10^{22}$	&	$2.07^{+0.22}_{-0.32}$	&	$39.64^{+0.09}_{-0.08}$	&	41.75/37	\\
			&23484	&	976	&	*$3.78\times 10^{20}$	&	$1.40^{+0.12}_{-0.11}$	&	$39.86^{+0.02}_{-0.02}$	&	59.47/53	&	$0.49^{+0.12}_{-0.14}\times 10^{22}$	&	$2.19^{+0.10}_{-0.22}$	&	$39.84^{+0.06}_{-0.07}$	&	58.45/53	\\
			&23485	&	1219	&	*$3.78\times 10^{20}$		&	$1.30^{+0.08}_{-0.08}$ 	&	$39.91^{+0.01}_{-0.02}$	&	52.63/63	&	$0.32^{+0.13}_{-0.11}\times 10^{22}$	&	$2.07^{+0.11}_{-0.10}$ 	&	$39.83^{+0.05}_{-0.05}$	&	70.32/63	\\
			&23486	&	624	&	*$3.78\times 10^{20}$			&	$1.13^{+0.13}_{-0.12}$	&	$39.62^{+0.03}_{-0.02}$	&	37.01/34	&	$0.14^{+0.21}_{-0.12}\times 10^{22}$	&	$2.07^{+0.33}_{-0.21}$	&	$39.51^{+0.13}_{-0.06}$	&	25.54/34	\\
			&23487	&	431	&		*$3.78\times 10^{20}$		&	$1.29^{+0.21}_{-0.18}$	&	$39.54^{+0.04}_{-0.04}$	&	22.57/22	&	$0.04^{+0.22}_{-0.04}\times 10^{22}$	&	$1.75^{+0.32}_{-0.22}$	&	$39.36^{+0.09}_{-0.01}$	&	16.21/22	\\
			&23490	&	610	&	*$3.78\times 10^{20}$		&	$1.23^{+0.14}_{-0.12}$	&	$39.63^{+0.03}_{-0.03}$	&	38.10/31	&	$0.30^{+0.18}_{-0.21}\times 10^{22}$	&	$2.19^{+0.22}_{-0.22}$	&	$39.58^{+0.10}_{-0.09}$	&	31.91/31	\\
			&23491	&	462	&	*$3.78\times 10^{20}$		&	$1.30^{+0.18}_{-0.13}$	&	$39.50^{+0.03}_{-0.04}$	&	33.98/24	&	$0.42^{+0.52}_{-0.38}\times 10^{22}$	&	$2.07^{+0.55}_{-0.32}$	&	$39.43^{+0.27}_{-0.13}$	&	29.87/24	\\

\botrule	
\end{tabular*}
{\scriptsize Note:- * hydrogen column density frozen to the average galactic value }
\end{sidewaystable*}	
\clearpage

\subsection{NGC 3079}\label{subsec1}
 Three point sources, namely CXOU J100202.9+553859 (X-1), CXOU J100205.4+554258 (X-2), and CXOU J100201.0+553648 (X-3), were identified in two observations (Obs. IDs 19307, 20947). The spectra of all three sources were well-fitted by both models. In these observations, each of the three sources exhibited an estimated X-ray luminosity ($L_X$) $\gtrsim 10^{39} $erg s$^{-1}$, placing them in the Ultraluminous X-ray Sources (ULXs) category. Furthermore, these sources manifested a hard spectral state with a power-law photon index ($\Gamma)  \sim $  2 (within error limits) when modeled with the powerlaw model. However, when the spectra of these three sources were interpreted using the disk blackbody model, their inner disk temperature (kT$_{in}$) $\gtrsim$ 1 keV in both observations. This estimated inner disk temperature far surpasses expectations from Eddington-limited blackbody accretion disks. Consequently, the radiative mechanism of these spectrally hard ULXs is proposed to involve the inverse Comptonization of soft photons within the hot plasma near the inner accretion disk, as suggested by previous studies \cite{rob2007, muk2015, jit2017}.

\subsection{PGC 032873}\label{subsec3}
	A source named CXOU J105604.8+421857 (X-4) was detected in all the four observations (ObsIDs- 21377, 21378, 22061, 22101) of the year 2019. Among the four observations, the spectrum of X-4, which recorded the highest count at approximately 226 (ObsID 21377), is effectively described by both models. However, in the remaining three observations, the spectral fit involves a limited number of degrees of freedom, roughly around 3 or 4 only. Consequently, the derived spectral parameters from these latter three observations lack a high level of conclusiveness. In each of these observations, X-4 has been identified as an ELX with a bolometric X-ray luminosity estimated to be  $\sim 1.44 \times 10^{41}$ erg s$^{-1}$ using the disk blackbody model. However, it is noteworthy that, as estimated by the powerlaw model, the X-ray luminosity within the 0.3-8.0 keV energy range, across all four observations, consistently registers at a level $\gtrsim$ $8.13 \times 10^{40}$ erg s$^{-1}$ within the margin of error limits. Spectrally, X-4 exhibits a consistently hard nature, characterized by a power-law photon index ($\Gamma$) of approximately 1.7 in all four observations, considering error limits. Moreover, if we interpret the spectra of X-4 in the context of disk accretion, the inner disk temperature (kT$_{in}$) is approximately 1.5 keV across all observations, accounting for error limits also. It's worth noting that this inner disk temperature of around 1.5 keV is notably higher than what would be anticipated from Eddington-limited blackbody accretion disks. Roberts (2007) \cite{rob2007}, Mukherjee et al. (2017) \cite{muk2015}, and Jithesh et al. (2017) \cite{jit2017} have previously reported that the observed hard emission from Ultraluminous X-ray Sources (ULXs) could be attributed to inverse Compton scattering of seed photons originating from the inner disc within a hot corona. Therefore, it seems plausible that the pronouncedly hard spectra exhibited by the ELX, X-4, may indeed be a result of inverse Comptonization of soft photons occurring in the vicinity of the accretion disk's hot corona.
	 
\subsection{NGC 1316}\label{subsec4}
Three point sources viz. CXOUJ032251.2-370950(X-5), CXOUJ032240.4-371640.6 (X-6) \& CXOUJ032240.8-371224 (X-7) with net counts $\ge$ 100 were detected in the different {\it Chandra} observations of NGC 1316 we have considered here for the present study. In both the observations-ObsID 20341 \& 22187, X-5 was found to be spectrally soft with an inner disk temperature, kT$_{in}$ $\sim $ 0.5 keV as estimated by disk blackbody model \& powerlaw photon index $\Gamma \sim 3$ as explained by powerlaw model. Nevertheless, it was observed that the spectra of X-5 exhibited a marginally better fit with the disk blackbody model in ObsID 20341. If the emission is interpreted as disk emission, the relation- disk blackbody normalization =  (R$_{in}/D_{10})^2$ cosi, can be employed. Here, R$_{in}$ is expressed in  kilometers, D$_{10}$ is the distance to the source in units of 10 kpc , and 'i' denotes the angle of the disk ( where i = 0 refers to face-on). For a point source like X-5 at distance, D $\sim$ 17 Mpc, we approximate the viewing angle, $cos\textit{i} \approx 0.5$ and taking the color factor, f $=$ 1.7 \cite{dev2008}, the color-corrected inner disk radius ($f^2 \times $R$_{in}$) can be estimated. 
Usually the mass of the compact object residing in ULXs is indirectly approximated using the disk blackbody model. Geometrically thin accretion disks are commonly believed to have an inner edge located near the innermost stable circular orbit (ISCO), which is approximately 3 times the Schwarzschild radius for a Schwarzschild black hole and around 0.5-4.5 times the Schwarzschild radius for a spinning Kerr black hole \cite{ori2000}. So, for a rough estimation of the black hole mass in point sources like ULXs at distances of a few megaparsecs, we assume the inner-disk radius, $R_{in} \sim$ 10 $GM/c^2$ (where G is the Universal Gravitational Constant, M is the mass of the Object , c is the speed of light in vacuum). Then, by using the color corrected inner disk radius, the mass of the compact object harbored by the corresponding ULX is estimated. Thus, the spectral parameters of X-5 with an estimated bolometric luminosity, $L_X \sim 3.26 \times 10^{39}$ erg s$^{-1} $ is consistent with a black hole of mass, $ M_{BH} \sim 137.86^{+66.62}_{-47.41} M_\odot $ which is accreting at $  \sim 0.19 $ times its Eddington limit.  The spectra of X-6 is slightly better fit by powerlaw model and that of X-7 is equally explained by both the models. Both the sources X-6 and X-7 are ULXs with their minimum luminosities $ > 10^{39}$ erg s$^{-1} $.  Both X-6 as well as X-7 are spectrally hard with their estimated inner disk temperature, kT$_{in}$ $\gtrsim$ 1 keV or the powerlaw photon index, $\Gamma \sim 2$ in all the observations. Analogous to the report of \cite{rob2007,muk2015,jit2017} for hard emission from ULXs,  the very hard spectra of X-6 and X-7 may also be due to inverse comptonization of seed photons in the hot corona near the inner accretion disk.
	
\subsection{NGC 4472}\label{subsec5}
A single point source  named CXOU J122923.8+075359 (X-8) was detected in three {\it Chandra} observations of the years- 2019 (ObsId 21647), 2020 (ObsId 21648) \& 2021 (ObsID 21649). The spectra of X-8 seem to be equally well fitted by both the considered models in the year 2019 \& 2020 observations, however in the year 2021 observation its spectrum is slightly better explained by the powerlaw model. In all the three epochs, X-8 is found to be spectrally hard with its powerlaw photon index $\Gamma \sim 1.41 - 1.75$. However, from the disk blackbody model, the inner disk temperature of X-8 is seen to become relatively softer in the subsequent years with kT$_{in}$ $\sim$ 1.38 keV (ObsId. 21648 of the year 2020)  and kT$_{in}$ $\sim$ 1.53 keV (ObsId.21649 of the year 2021) as compared to its relatively harder state with kT$_{in}$ $\sim$ 1.89 keV in the year 2019 observation (Obs. Id. 21647). Since such hard spectrum with kT$_{in}$ $\gtrsim$ 1.0 keV originates from the hot flow of energetic particles near the inner accretion disk, the relatively softening of the hard spectra of X-8  in its different epoch is likely due to different amount of Compton cooling acting on the accretion disk corona as it is irradiated by softer radiation from the central compact source. The estimated  bolometric luminosity of X-8 in all the three epoch remain in the ELXs range with $ L_X > 10^{40}$ erg s$^{-1}$ even within error limits.

\subsection{IRAS 18293-3413}\label{subsec6}
	Two sources, CXOU J183241.9-341625 (X-9) and CXOU J183224.7-340917 (X-10) were detected from the {\it Chandra} observation- ObsId 21379. The spectra of both X-9 \& X-10 are equally well explained by both the considered empirical models.  X-9 is a possible Hyperluminous X-ray source (HLX) with its minimum luminosity $> 10^{41}$ erg s$^{-1}$ while X-10 is an Extremely luminous X-ray source (ELX) with its minimum luminosity $> 10^{40}$ erg s$^{-1}$ . The spectra of both these two sources are hard with their powerlaw photon index, $\Gamma \sim 2.0$, within error limits. If tried to explain by disk blackbody model, both X-9 \& X-10 have an inner disk temperature around 1.5 keV and above, which actually is very much higher than those expected from Eddington limited blackbody accretion disks and hence the radiative mechanism of these two spectrally hard \& extremely luminous sources may indeed be inverse comptonization of soft photons near the extreme environment of the accretion disk.
\subsection{NGC 1600}\label{subsec7}
Two point source CXOU J043147.4-050213 (X-11) and CXOU J043139.7-050154 (X-12) were observed from two different {\it Chandra} observations (Obs. Id.21375, 22878). The spectra of both the sources are preferentially well explained by powerlaw model. Both X-11 and X-12 are found to be ELXs with their luminosity $> 10^{40}$ erg s$^{-1}$. The spectra of both these sources are found to be hard with powerlaw photon index ($\Gamma) $ around 1.8. The radiative mechanism of such hard and extremely luminous X-ray sources (ELXs) may be due to inverse comptonisation of soft photons.
\subsection{NGC 7793}\label{subsec8}
In the {\it Chandra} observation (ObsID 23266) of NGC 7793, four X-ray sources viz. CXOU J235806.7-323756 (X-13), CXOU J235751.0-323726 (X-14), CXOU J235743.8-323634 (X-15) \& CXOU J235746.7-323607 (X-16) were detected.  The spectra of X-13, X-15 \& X-16 are equally well explained by both the powerlaw model as well as the disk blackbody model and their estimated luminosities by both the models lie in the range of X-ray binaries (L$_X \sim 10^{38}$ erg s$^{-1}$). These three X-ray binaries are all spectrally hard with their powerlaw photon index, $\Gamma \lesssim 2$ within error limits, which actually may be attributed as due to inverse comptonization of soft photons.    X-14 is found to have a very hard spectrum with its inner disk temperature, kT$_{in} \sim 2.0 $ keV and its estimated bolometric luminosity, L$_X \sim 1.09 \times 10^{39}$ erg s$^{-1}$ which is in the ULX range, however the (0.3-8.0) kev range estimates its luminosity in the X-ray binary range $\sim 4.46 \times 10^{38}$erg s$^{-1}$. 
\subsection{NGC 4214}\label{subsec9}
Three X-ray sources namely CXOU J121538.3+361920 (X-17), CXOU J121538.2+361944 (X-18) \& CXOU J121541.4+362114 (X-19) were detected from the {\it Chandra} observation - ObsId 22372 of the year 2020. All these three sources are equally well explained by both the models used here. \cite{dewi2006cxou} \cite{ghosh2006discovery}, using XMM-Newton, ROSAT \& {\it Chandra} data reported X-17 as the brightest X-ray source in NGC 4214 with X-ray luminosity $\sim 0.7 \times 10^{39} $erg s$^{-1} $ and \cite{binder2015chandra} using {\it Chandra} observational data of the year 2001 and 2004 also reported X-17 to be an XRB which is likely composed of a slightly evolved He-burning donor star and a lower-mass compact companion which may either be a neutron star or a small mass blackhole. In the present work with {\it Chandra} obervational data of the year 2020 also, X-19 is found to be a spectrally hard X-ray binary with X-ray luminosity, $L_X \sim 4.26 \times 10^{38}$ erg s$^{-1} $ and its powerlaw photon index, $\Gamma \approx 1.86^{+0.11}_{-0.11}$. If, we try to explain the spectra of X-17 by disk blackbody model and also assuming isotropic emission, then its spectral parameters are consistent with this X-ray binary to host a stellar mass blackhole accreting at super-Eddington rate. Likewise X-18 and X-19 are also found to be X-ray binaries which are also spectrally hard, $\Gamma \sim$ 2.0 for X-18 \& 1.6 for X-19, within error limits. However, X-18 \& X-19 under assumption of isotropic emission and their spectra explained by disk blackbody model, their spectral parameters estimates for them to harbor stellar mass black holes accreting at sub-Eddington rate.

\subsection{NGC 4485/90}\label{subsec10}
	A total of 8 sources namely CXOUJ123043.1+413820 (X-20),CXOUJ123036.2+413839  (X-21),CXOUJ123035+413847 (X-22), CXOUJ123030.7+413912 (X-23),CXOUJ123032.2+413919  (X-24),CXOUJ123029.4+413928   (X-25), CXOUJ123031.7+414142 (X-26) and CXOUJ123030.5+414142 (X-27) were observed from  {\it Chandra's} eight observations (ObsIds - 23482, 23483, 23484, 23485, 23486, 23487, 23490, 23491). These eight sources are estimated to be ULXs as their estimated bolometric luminosities are all $\gtrsim10^{39}$ erg s$^{-1}$.
	
	The spectra of CXOUJ123043.1+413820 (X-20) are equally well explained by both the models in all the different observations considered here. X-20 is spectrally hard with an inner disk temperature, kT$_{in}$ ranging between 1.18-1.81 keV as estimated by disk blackbody model \& powerlaw photon index $\Gamma \sim 2$ as explained by powerlaw model. Earlier work by \cite{gladstone2009ultraluminous}, using {\it Chandra} observational data of the year 2004 (ObsID 4725 \& 4726) and XMM-Newton observational data of 2002, has reported the spectra of this source hardening consistently with increase in luminosity. However with the recent {\it Chandra} data considered for the present study, X-20 becomes most luminous even reaching the ELX range with its minimum luminosity greater than $ 10^{40} $erg s$^{-1}$  in its softest state with kT$_{in} \sim 1.18^{+0.08}_{-0.10}$/$\Gamma \sim 2.41^{+10}_{-11}$ in the ObsID 23486. While in its hardest state with kT$_{in} \sim 1.76^{+0.35}_{-0.23}$/$\Gamma \sim 1.86^{+21}_{-22}$ (ObsID 23483) its luminosity was $\sim 4.78 \times 10^{39}$ erg s$^{-1}$. Despite the persistent hardness observed in the spectra of X-20 across all considered observations, there are intermittent patterns of both hardening and relative softening. Moreover, these variations are accompanied by slight changes in luminosity over timescales of months. This dynamic behavior could be attributed to different amounts of Compton heating/cooling influencing the accretion disk corona. These effects are likely a consequence of varying irradiation by harder or softer radiation from the central compact sources, coupled with fluctuations in accretion rates.
	
   In the present study, CXOUJ123036.2+413839 (X-21) is detected to be a ULX with its bolometric luminosity$ > 10^{39}$erg s$^{-1}$ in all the epochs of observations considered here. In all the observations, the spectra of X-21 is also equally well explained by both the models. X-21 is found to be spectrally hard with an inner disk temperature kT$_{in}$ ranging from 0.76 keV (ObsId. 23482) to 1.41 keV in (ObsId. 23486) as explained by disk blackbody model and powerlaw photon index $\Gamma \gtrsim 2$ as explained by powerlaw model. Using Chandra observational data of the year 2004 (ObsID 4725 \& 4726) and XMM-Newton observational data of 2002,  \cite{gladstone2009ultraluminous} reported this source to get hardening with the increase in its luminosity. In the present study, with the recent most {\it Chandra } data considered here we find that the source spectra gets harder as its luminosity increases - it has the least luminosity L$_X \sim 4.89 \times 10^{39}$ erg s$^{-1}$  in its softest state with kT$_{in} \sim$ 0.76 keV (ObsID 23482) while in its one of the hardest state with kT$_{in} \sim 1.39^{+0.12}_{-0.14}$ (ObsID 23487) it is extremely luminous with a bolometric luminosity, L$_X \sim 1.34 \times 10^{40}$ erg s$^{-1}$ which has even entered to the ELX range. For X-21 also, similar to the observed behavior in X-20, the intermittent spectral hardening/softening, coupled with varying luminosities, may be ascribed to fluctuations in Compton heating/cooling processes accompanied with changing accretion dynamics.
	
	CXOUJ123035.1+413847 (X-22) is detected significantly only in one observation ( ObsId 23484). The spectra of X-22 is better fit by the disk blackbody model with its $cstat/dof \sim 0.89 $ as compared to the value, $cstat/dof \sim 0.50 $ of the powerlaw model. This source is found to be a hard ULX with its inner disk temperature, kT$_{in}$ around 1.64 keV and its bolometric luminosity, $L_X\sim 1.90 \times 10^{39}$ erg s$^{-1}$. 
	
	The source CXOUJ123030.7+413912 (X-23) is detected in all the epoch of observations of NGC 4485/90 considered for the present study. In all its obervations, the spectra of X-23 was found to be equally well expalined by both the models, however in one observation with ObsId. 23485, the spectral fit with either of the models were very poor ($cstat/dof \sim 0.3 $). This may be because of low number of bins to fit. In all the other observations, the luminosity was found to be consistently in the ULX range with L$_X \sim 7.94 \times 10^{39}$ \ erg s$^{-1}$ within the uncertainty limits. Despite the consistency in its bolometric luminosity its spectral parameters such as inner disk temperature/spectral slope seems to vary slightly. With Chandra observational data of the year 2004 (ObsID 4725 \& 4726) and XMM-Newton observational data of 2002, \cite{gladstone2009ultraluminous} also found that X-23 varies its spectral slope with an almost constant observed luminosity, which agrees well with the current finding with the recentmost {\it Chandra} observational data. 
	
	CXOUJ123032.2+413919 (X-24) was also detected in all the epoch of observations of NGC 4485/90 considered for the present work. The spectra of this source is also equally well explained by both the models in almost all its epochs of observations. In some of its observations (ObsID 23482, 23483, 23484 \& 23485), the spectral state of this source seem to soften while its luminosity remains nearly constant. However in some of the observations (ObsID 23486, 23487, 23490 \& 23491) there were slight variations in the luminosity with variations in the spectral state, but there were no perfect correlation between these variations. A significant decrease in the bolometric luminosity from $1.02 \times 10^{40}$ erg s$^{-1}$ (ObsId 23486) to $4.67 \times 10^{39}$ erg s$^{-1}$ (ObsId 23487) was observed in just a period of nearly one month with a slight softening of the spectra. Such slight spectral softening, coupled with decreased luminosity by a factor of around 2 in timescale of $\sim $1 month may be due to altered accretion rate along with Compton cooling due to irradiation of the accretion disk corona by  softer radiations. \cite{gladstone2009ultraluminous} reports X-24 to be better fit by disk blackbody model with no substantial change in luminosity and disk temperature in the year 2000 and 2004 {\it Chandra} \& 2002 XMM-Newton observations of their study. 
	
	CXOUJ123029.4+413928 (X-25) is detected to be a ULX ($L_X > 10^{39}$ erg s$^{-1}$) in all its observed epochs here. \cite{gladstone2009ultraluminous} identified X-25 as a possible supernova remnant candidate and \cite{vazquez2007constraints} reported this source being coincident with a radio source FIRST J123029.4 + 413927.The spectra of X-25 is equally well explained by both the models except for few instances where powerlaw model is slightly prefered ( ObsId 23484). There were variations in the spectral parameters with luminosity, however the variation is more significant in case of powerlaw model where the spectral slope ranges between ( 1.63 - 3.61). Also it is seen that the softest state with $\Gamma \sim 3.61$ (ObsId 23484) corresponds to the most luminous state ($L_X \sim 2.51 \times 10^{40}$ erg s$^{-1}$) while the hardest state with $\Gamma \sim 1.63$ ( ObsId 23490) corresponds to the least luminous state ($L_X \sim 2.18 \times 10^{39}$ erg s$^{-1}$). This observed spectral state transition from a relative soft state with high luminosity to a hard state with low luminosity is very much like the soft to hard spectral state transition seen in many X-ray binaries. Such spectral state transition from high-soft to low-hard state signal the change between accretion modes - disk accretion to a hot advection-dominated flow (ADAF) \cite{liu2005}. Furthermore, the substantial change in luminosity by a factor of $\sim$ 11 within a period of around 6 months suggests a drastic alteration in the accretion rate. Such significant variations in luminosity and spectral states can be indicative of dynamic changes in the accretion processes occurring in the system. Using two {\it Chandra} epoch of observations of 2004 \cite{fridriksson2008long} also reported X-25 to have a fall of 30 \% in its luminosity within a period of four months which is more typical of an X-ray binary.  
				
The sources CXOUJ123031.7+414142 (X-26) and CXOUJ123030.5+414142 (X-27) are detected to be Ultraluminous X-ray sources with their bolometeric luminosity (L$_X) > 10^{39}$ erg s$^{-1}$ in all its observed epochs considered here. The spectra of both these sources are found to be equally well explained by both the models considered. Source X-26 is spectrally hard with its inner disk temperature, kT$_{in} $ ranging between $0.82^{+0.19}_{-0.14} - 1.58^{+0.20}_{-0.15}$ keV as explained by disk blackbody model or powerlaw photon index $\Gamma \sim $ 2 (within error limits) if explained by powerlaw model. Considering all the epochs here it is clearly seen that X-26 has a trend of increasing its luminosity with the hardening of its spectra. In its softest state with kT$_{in} \sim 0.82$ keV (ObsID 23483) its bolometric luminosity is $ \sim 1.28 \times 10^{39}$ erg s$^{-1}$ which after a period of nearly 3 months in the observation with ObsID 23486 the spectra is in the hardest state with kT$_{in} \sim 1.58$ thereby its bolometric luminosity being increased to $ \sim 6.76 \times 10^{39}$ erg s$^{-1}$, i,e. by a factor $\sim$ 5.28. This may actually be due to enhanced accretion rate along with compton heating in the accretion disk corona. For the source X-27, overall, the spectra is hard with its inner disk temperature, kT$_{in} > $ 1 keV if explained by the disk blackbody model, and if explained by powerlaw model, its photon index, $\Gamma \sim 2.0$ within error limits. In most of the epochs here, the luminosity of X-27 seem to increase with spectral hardening. For the same source, similar result of hardening the spectra with an increase in luminosity were also reported by \cite{gladstone2009ultraluminous} by using only two epochs of {\it Chandra} observation in the year 2004. In agreement to these findings, \cite{avdan2019optical} also reported X-27 to have a  slight softening of the spectra with decrease in luminosity in the two epochs of XMM Newton observations in the year 2008.

\section{Conclusion}\label{sec4}
We present the results of the spectral analysis of point X-ray sources in a sample of nine galaxies recently observed by {\it Chandra} ACIS-S during the year 2018-2021. 27 point X-ray sources were found to have net source counts $\ge$ 100 and hence considered for the spectral analysis. The spectra of the 27 sources were fitted using two empirical models - an absorbed disk blackbody model and an absorbed powerlaw model. From the estimated X-ray luminosity, the 27 X-ray sources are categorized as 6 XRBs and 21 ULXs. Almost all the sources were equally well explained by both the models, except for some sources in some of their particular observations, the spectra prefer one model slightly over the other.  All the six XRBs - X-13, X-15, X-16, X-17, X-18 and X-19 were found to be spectrally hard  with their powerlaw photon index, $\Gamma$ ranging between $1.52-2.29$ . The radiative mechanism of these  spectrally hard X-ray binaries may indeed be inverse comptonization of soft photons near the extreme environment of the accretion disk. Out of the 21 ULXs detected, only the ULX -  CXOUJ032251.2-370950 (X-5) was found to be spectrally soft with an inner disk temperature, kT$_{in}$ $\sim $ 0.5 keV as estimated by disk blackbody model \& powerlaw photon index $\Gamma \sim 3$ as explained by powerlaw model. The spectra of X-5 was found to be slightly better fit by disk blackbody model in the ObsID 20341. Thus, if the emission is considered to be disk emission with the inner disk radius, $R_{in} \sim $10GM/c$^2 $ , the spectral parameters of X-5 with an estimated bolometric luminosity, $L_X \sim 3.26 \times 10^{39}$ erg s$^{-1} $ is consistent with a black hole of mass, $ M_{BH} \sim 137.86^{+66.62}_{-47.41} M_\odot $ which is accreting at $  \sim 0.19 $ times its Eddington limit. All the other remaining ULXs are all spectrally hard and hence inverse comptonization of seed photons in the hot plasma near the inner accretion disk may be attributed as their emission mechanism.\\
	Out of the 21 ULXs, 8 ULXs - CXOUJ105604.8+421857(X-4), CXOUJ122923.8+075359(X-8), CXOUJ183241.9-3416259(X-9), CXOUJ183224.7-340917(X-10), CXOUJ043147.4-050213(X-11), CXOUJ043139.7-050154(X-12), CXOUJ123043.1+413820(X-20) and CXOUJ123036.2+413839(X-21) are found to have even the lower limit of their luminosity $> 10^{40}$ erg s$^{-1}$ and hence they are ELXs. In some of the ULXs/ELXs softening/hardening of spectra with or without changes in their luminosities were also observed. X-8 is a hard ELX with consistent luminosity while its spectra was found to become relatively softer in the subsequent years - 2020 observation (ObsId. 21648)  and 2021 observation (ObsId.21649) as compared to its relatively harder state in the year 2019 observation (Obs. Id. 21647). X-20 is a hard ULX which has become most luminous even reaching the ELX range ( $L_X \gtrsim 10^{40}$ erg s$^{-1}$)  in its softest state (ObsID 23486), while its luminosity was estimated the least in its hardest state (ObsID 23483). Similar trend follows in case of X-25 also, where it becomes the most luminous in its softest state (ObsID 23484). However in case of X-21 and X-26, the spectra gets harder with increase in luminosity. The hardest state of X-21 (ObsID 23487) correspons to the  most luminous state with a bolometric luminosity, L$_X \sim 1.34 \times 10^{40}$ erg s$^{-1}$ which has even entered the ELX regime, likewise for X-26 also its most luminous state corresponds to the hardest state (ObsID 23486). Such spectral hardening/softening, coupled with varying luminosities, may be ascribed to fluctuations in Compton heating/cooling processes accompanied with changing accretion dynamics. \\
	In future, a comparative study of these X-ray sources by using data from other missions such as XMM-Newton, NuSTAR, AstroSAT etc., wherever available, will help to understand the physical nature of the sources in more details.

\bmhead{Acknowledgments}

The authors would like to thank the {\it Chandra} X-ray Center Archive for its resources of data that have been used in the present work.

\end{document}